# Bulk Nanostructured Zirconia Ceramics with High Hardness and Toughness via Integration of High-Pressure Torsion and Spark Plasma Sintering

Kaveh Edalati[1,2,]*, Koji Morita[3], Shivam Dangwal[1,2] and Zenji Horita[4-6]

[1] WPI, International Institute for Carbon Neutral Energy Research (WPI-I2CNER), Kyushu University, Fukuoka 819-0395, Japan
[2] Department of Automotive Science, Graduate School of Integrated Frontier Sciences, Kyushu University, Fukuoka 819-0395, Japan
[3] National Institute for Materials Science, Tsukuba, Ibaraki 316-8511, Japan
[4] Graduate School of Engineering, Kyushu Institute of Technology, Kitakyushu 804-8550, Japan
[5] Magnesium Research Center, Kumamoto University, Kumamoto 860-8555, Japan
[6] Synchrotron Light Application Center, Saga University, Saga 840-8502, Japan

Developing nanostructured bulk ceramics is a major challenge when conventional high-temperature sintering is employed for consolidation. In the current investigation, yttria-stabilized zirconia (YSZ) with a composition of $ZrO_2$ - 3 mol% $Y_2O_3$ is first treated using high-pressure torsion (HPT) and further consolidated using spark plasma sintering (SPS) to produce a nanostructured bulk sample. The material demonstrates phase transformations from tetragonal to dislocation-decorated monoclinic by HPT and reversely transforms to the tetragonal phase after the SPS process while maintaining a mean grain size of 80 nm and large numbers of dislocations. The consolidated ceramic exhibits a density of 6.07 g/cm$^3$ (99% relative density) with a high hardness of 1500 Hv, which is reasonably consistent with the prediction of the Hall-Petch relationship. Examination of the indented areas during the hardness test confirms the absence of cracks, indicating good fracture toughness ($K_{IC}$) because of the presence of dislocations, while the sample processed only by SPS and without HPT processing forms numerous cracks by indentation and exhibits low $K_{IC}$.

***Keywords:*** ceramic materials; ultrafine-grained (UFG) materials; severe plastic deformation (SPD); ductility; Hall-Petch relation

*Corresponding author (E-mail: kaveh.edalati@kyudai.jp; Tel/Fax: +81 92 802 6744)

# 1. Introduction

Yttria-stabilized zirconia (YSZ) ceramic is widely recognized for its exceptional mechanical properties, thermal stability, and ionic conductivity, making it a crucial material in various engineering applications, including thermal barrier coatings, fuel cells, dental sectors and structural ceramics [1-4]. YSZ primarily exists in the tetragonal or cubic crystal forms at ambient temperature when stabilized with yttria [1]. While YSZ exhibits transformation toughening through stress-mediated tetragonal-to-monoclinic or cubic-to-monoclinic phase transitions [5], further improvement of its toughness is necessary to enhance its reliability in structural applications. Enhancing the toughness of YSZ has been a significant challenge for several decades like any other ceramic materials, particularly because of its application as a biomaterial [6,7].

One promising approach to improving the mechanical performance of YSZ is to refine its grain structure to the nanoscale [8,9]. Nanostructured ceramics often exhibit enhanced hardness because of the Hall-Petch effect, where reduced grain size directs to increased grain boundary strengthening [10]. However, a major concern with nanocrystalline ceramics is their inherent brittleness, as grain boundaries can act as crack initiation sites [11]. Recent studies suggest that introducing a high density of dislocations into ceramics can enhance their toughness by hindering premature microcrack propagation [12,13]. Therefore, a material that is both nanostructured and dislocation-rich has the potential to exhibit not only high hardness but also improved toughness, overcoming the typical trade-off between strength and toughness in ceramics. To achieve such a microstructure, advanced processing techniques are required to refine grains while introducing and stabilizing hard-to-form dislocations in ceramics [14].

Severe plastic deformation processes [15,16], like high-pressure torsion (HPT) [17,18], have been extensively studied for their ability to generate a high concentration of defects and refine microstructures in metals and metallic alloys. Although ceramics are generally brittle and difficult to process using severe plastic deformation, recent research has shown that HPT can induce phase transformations and grain refinement in various ceramics [19-21], making it a promising technique for enhancing mechanical properties. However, HPT alone does not achieve full densification in ceramics, and an additional consolidation step is required, as reported earlier for $Al_2O_3$ [22], $BaTiO_3$ [23] and WC [24] ceramics. Spark plasma sintering (SPS) has emerged as an advanced sintering process that achieves fast consolidation at lower temperatures and shorter sintering times, effectively minimizing grain growth and dislocation recovery [25,26]. By combining HPT with SPS, it may be possible to produce fully dense, nanostructured YSZ ceramics with an enhanced density of dislocations, leading to both increased hardness and improved toughness. The synergy between these two techniques can potentially provide a novel pathway for optimizing the mechanical properties of bulk ceramics by grain reduction to the nanometer range, defect engineering and controlled densification.

In the present research, the combination of HPT and SPS is applied to process $ZrO_2$ - 3 mol% $Y_2O_3$ ceramics for the first time, aiming to achieve a nanograined bulk material with enhanced hardness, density and toughness. The phase evolution, microstructural features, and mechanical property evolvement of the processed material are analyzed through X-ray diffraction (XRD), Raman spectroscopy, scanning electron microscopy (SEM), transmission electron microscopy (TEM), hardness testing and density measurements. A comparison of the variations of hardness and density demonstrates the effectiveness of the approach in enhancing both hardness and toughness.

## 2. Experimental Procedure

The commercial partially stabilized $ZrO_2$ - 3 mol% $Y_2O_3$ nanopowder was used in this study. Approximately 500 mg of the powder was put on a circular hole located in the lower anvil of an HPT facility. The powder was first pressed employing an applied pressure of 6 GPa between the lower and upper anvils of HPT, and after the stabilization of the pressure, it was torsionally strained through turning the lower HPT anvil against the upper anvil, as described in detail elsewhere [17,18]. The HPT process was carried out under ambient temperature with a rate of 1 turn per minute. About 20 samples, which had disc shapes with a 5 mm radius and 0.8 mm height, were prepared by HPT. All HPT-treated discs were crushed and mixed together using a mortar and pestle and further processed by spark plasma sintering (SPS) using a facility described elsewhere [27]. SPS was conducted using a graphite die under vacuum conditions to obtain dense bulk disc-shaped samples with 5 mm radius and 1 mm height. For SPS, the applied pressure was 80 MPa, the rate of heating was 100 K/min, the maximum temperature was 1473 K, the maintaining period at the highest temperature was 3 min, and heating was controlled by direct current pulses. In addition to disc samples, dog-bone-type tensile specimens were also prepared by SPS and HPT followed by SPS. However, the tensile specimens could not be consolidated appropriately because of their complex shape, and they broke in the elastic region during testing.

After consolidation via SPS, the disc-shaped samples were polished on both sides and examined by different methods, including XRD, Raman spectroscopy, SEM, TEM, density measurement according to ASTM B962 standard, and hardness testing according to ASTM C1327 standard. For comparison, the starting powders and the compacted powders after HPT treatment were also examined. To investigate phase transformations, XRD with Cu Kα radiation, 0.05º scanning step size and 1º/min scan speed, and Raman spectroscopy with a 532 nm laser source were used. For TEM observations, about 5 mg of the samples were crushed in ethanol using a mortar and pestle and poured on a copper grid with carbon film coverage. TEM was performed under 200 kV to take bright- and dark-field micrographs, record selected area electron diffraction (SAED), and examine nanostructural features using high-resolution images and relevant diffractograms achieved via fast Fourier transform (FFT) analysis. Density measurement was conducted using the Archimedes principles, employing a digital balance with 0.0001 g accuracy. Hardness was determined by the Vickers method, employing a force of 1 kgf for 15 s. The intended regions were examined by field-emission SEM and energy-dispersive X-ray spectroscopy under 15 kV. The samples before SEM analysis were placed in a plasma coating machine and were coated using gold to avoid electron charge-up.

## 3. Results

The appearance of pieces of samples consolidated by SPS and HPT followed by SPS is shown in Fig. 1. While the sample consolidated only by SPS has a bright color, the sample processed by HPT followed by SPS has a dark color. The dark color usually indicates the existence of crystal lattice imperfections like vacancies and dislocation-type defects in ceramic materials [27]. While such defective black ceramics have received attention in the field of photocatalysis [28-30], they can also be of interest for mechanical properties because one reason for the brittleness of ceramics is the difficulty in generating lattice defects. It has already been proved using electron spin resonance that vacancies are formed in YSZ by the HPT treatment [30], but part of such vacancies apparently survive during rapid consolidation by SPS.

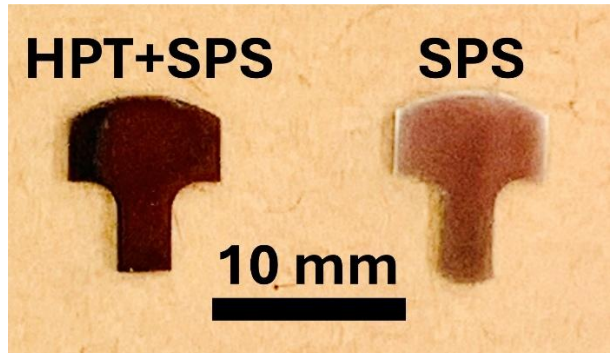

Figure 1. Appearance of yttria-stabilized zirconia samples consolidated by high-pressure torsion followed by spark plasma sintering (HPT+SPS) and only spark plasma sintering (SPS).

The examination of phase transformation by HPT and SPS is illustrated in the XRD patterns of Fig. 2a. The initial powder contains two phases: monoclinic (JCPDS Card No.: 01-070-8739), which is the thermodynamically stable phase of pure zirconia, and tetragonal (JCPDS Card No.: 01-079-1765), which is a high-temperature crystal form of pure zirconia that is stabilized at ambient temperature by the addition of yttria [1-5]. Examination of the fraction of phases using the method suggested in Ref. [31] suggests that the initial powder contains 66 vol% of tetragonal and 34 vol% of monoclinic. After HPT processing, the intensity of the monoclinic peaks increases, and the amount of the tetragonal phase reduces to 32 vol%. Besides the tetragonal-to-monoclinic transition, significant peak broadening appears for the two phases. Such peak broadening is due to crystallite size reduction and dislocation generation [32]. The generation of dislocations in ceramics by HPT, which is a direct effect of straining, has been reported in various ceramics [33]. Moreover, phase transformations are observed frequently in HPT-processed ceramics [19-21]. After SPS processing, the material transforms to 100 vol% of the tetragonal phase, as shown more clearly in Fig. 2b using Rietveld refinement. The sample processed by only SPS also exhibits 100% of the tetragonal phase with no difference in lattice parameters or XRD peak shift compared to the sample treated by integrating HPT and SPS. Raman spectra, compared to the reference data [34], as shown in Fig. 2c, also confirm the phase transformations to the monoclinic phase by the HPT treatment and to the tetragonal phase by the SPS treatment. It should be noted that the transition of the monoclinic structure to high-temperature tetragonal or cubic phases usually occurs during SPS [35].

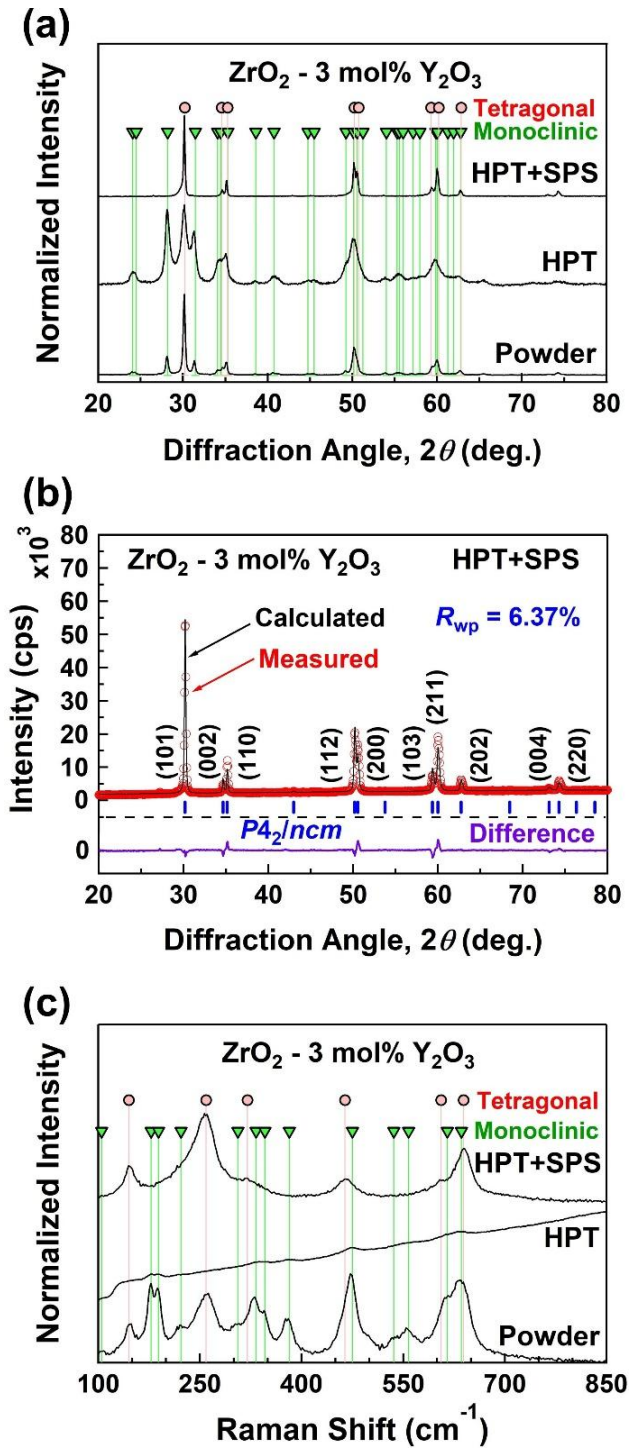

Figure 2. (a) X-ray diffraction profiles for yttria-stabilized zirconia including initial powder, sample treated by high-pressure torsion (HPT) and disc sintered using high-pressure torsion followed by spark plasma sintering (HPT+SPS), (b) Rietveld refinement for HPT+SPS-processed sample ($R_{wp}$: weighted profile $R$-factor), and (c) Raman spectra for initial powder and samples treated by HPT and HPT+SPS.

Microstructural examinations using TEM bright- and dark-field micrographs as well as SAED analysis are summarized in Fig. 3 for (a) initial powder, (b) HPT-processed sample and (c) HPT+SPS-processed sample. Examination of samples at higher magnifications is shown in Fig. 4 for (a) starting YSZ powder, (b, c) HPT-treated material and (d, e) HPT+SPS-processed sample. The initial powder contains nanoparticles with a mean size of 72 ± 26 nm. Both monoclinic and tetragonal phases can be found in these particles, as illustrated in Fig. 4a by a high-resolution image as well as FFT analysis. Following the HPT treatment, the grains are reduced to a mean size of 26 ± 14 nm, while grain boundaries become ambiguous in dark-field images. The decrease in the grain size by HPT can also be confirmed by the shape of the SAED profiles, which change to fully ring-shaped patterns. In the HPT-treated material, because of the diffraction contrast from strained nanograins, it is difficult to observe dislocations using bright- and dark-field micrographs; however, numerous dislocations can be observed in high-resolution images illustrated in Fig. 4b and 4c for the monoclinic and tetragonal phases, respectively. These dislocations were suggested to contribute to the acceleration of phase transformations in some studies [21,36]. After applying SPS, grain coarsening occurs to a size range of 37-126 nm measured for 40 grains, with an average grain size of 80 ± 44 nm, which is still at the nanometer level. The change in the SAED profile to a dotted pattern also suggests the occurrence of grain coarsening, a phenomenon that is controlled by lattice diffusion, grain boundary migration and/or dislocation climb during SPS [35]. Examination of these grains at higher magnifications, as illustrated in Fig. 4d and 4e, verifies the existence of numerous dislocations (see dark lines in grains) in the sample after SPS processing. These observations confirm that a bulk nanograined sample with a large fraction of dislocations could be produced by integrating HPT and SPS processes. These microstructural features indicate the high potential of the sample for enhanced mechanical properties [8-13].

The density of three different samples after processing by HPT followed by SPS is 6.07±0.09 g/cm³. This value is near the nominal density of the tetragonal crystal in $ZrO_2$ - 3 mol% $Y_2O_3$, which is usually 6.13 g/cm³ [37]. If the nominal density is considered 6.13 g/cm³, it can be concluded that a relative density as high as 99% is obtained by the combination of HPT and SPS, indicating enhanced consolidation by this integrated process. Earlier studies on the application of conventional sintering after HPT processing also reported good consolidation, but they did not succeed in achieving such a high relative density [22-24].

The variation of hardness across the disc diameter after processing by HPT followed by SPS is shown in Fig. 5 against the distance from the disc center. The material exhibits high homogeneity across the disc radius, with an average hardness of 1500 ± 70 Hv. This hardness is considered high for bulk samples of $ZrO_2$ - 3 mol% $Y_2O_3$, which typically show hardness levels in the range of 1150–1450 Hv [2,38,39]. It should be noted that the hardness of the sample consolidated only by SPS reaches 1450 ± 30 Hv, which is somewhat smaller than the hardness of the material processed by HPT and SPS. This enhancement in hardness is attributed to the presence of nanograins and dislocations [8-13].

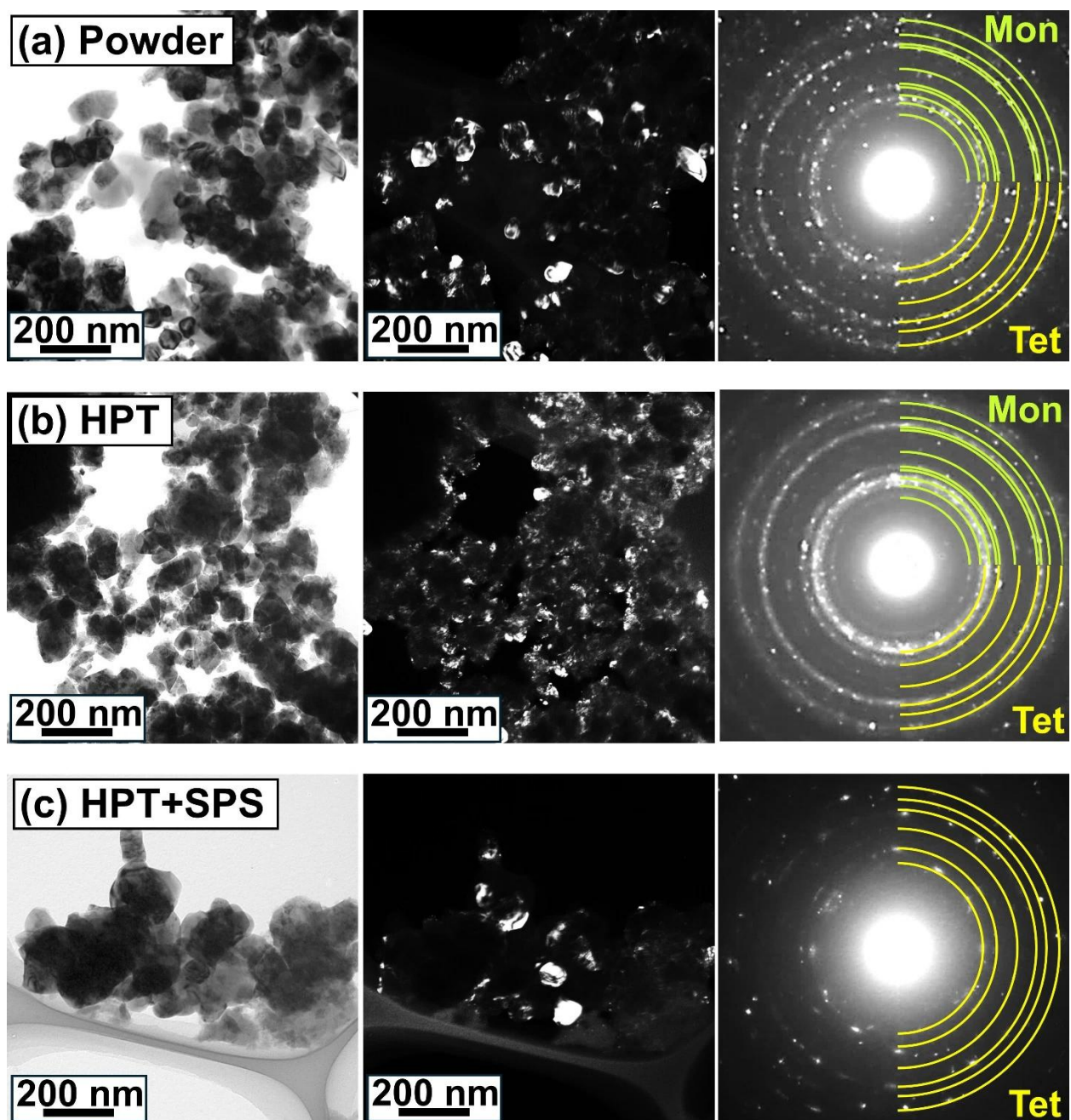

Figure 3. Bright- (right) and dark-field (center) micrographs and selected area electron diffraction (SEAD) profiles (left) obtained using transmission electron microscopy for yttria-stabilized zirconia including (a) initial powder, (b) sample treated using high-pressure torsion (HPT) and (c) disc sintered using high-pressure torsion followed by spark plasma sintering (HPT+SPS). Mon and Tet in SEAD patterns refer to monoclinic and tetragonal phases, respectively.

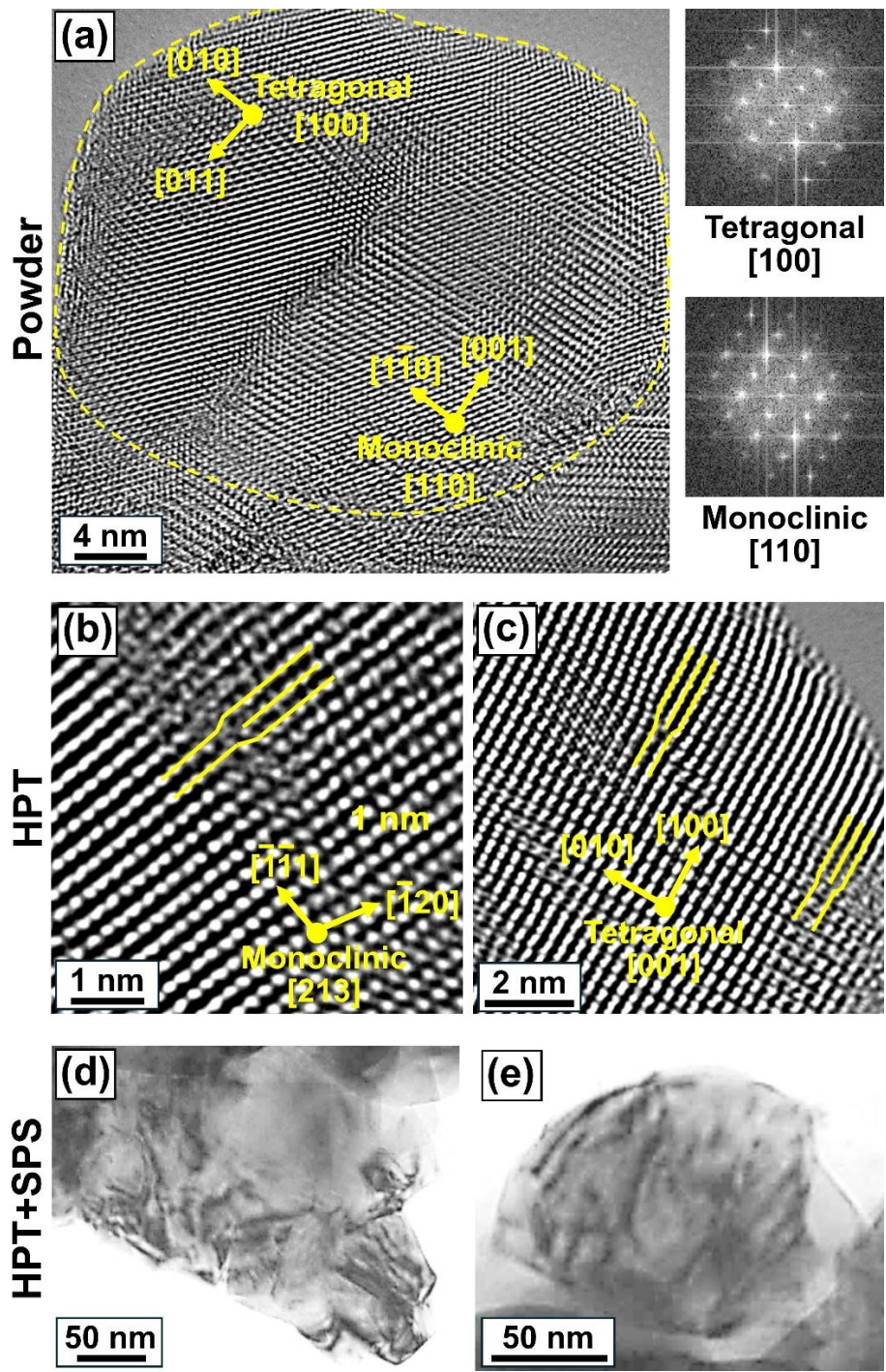

Figure 4. Magnified views of microstructure obtained using transmission electron microscopy for yttria-stabilized zirconia including (a) high-resolution lattice image (left) and relevant fast Fourier transform diffractograms (right) for initial powder, (b, c) lattice images of dislocations in material treated using high-pressure torsion (HPT) and (d, e) high-magnification bright-field images for disc sintered by high-pressure torsion followed by spark plasma sintering (HPT+SPS).

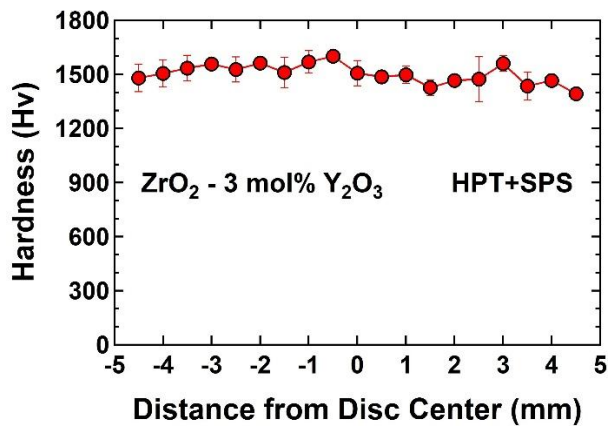

Figure 5. Variation of Vickers microhardness as a function of radial distance for yttria-stabilized zirconia disc sintered using high-pressure torsion followed by spark plasma sintering (HPT+SPS).

To evaluate the toughness of the material, tensile and indentation tests were performed. Tensile testing was not successful due to technical problems in the consolidation of specimens with complex shapes. However, the results of indentation testing are summarized in Fig. 6a for 10 indented regions. For comparison, indentation was also conducted on the sample consolidated only by SPS, and the results are included in Fig. 6b. For the sample treated using HPT followed by SPS, cracks are not observed at the edges of the indentation area, except for one tip at two indented areas. This confirms that the material exhibits high toughness despite its high hardness. In contrast, for the sample processed only by SPS, cracks are observed at all tips of the indented areas in Fig. 6b. Closer examination of the indented regions using field-emission SEM and EDS, as shown in Fig. 7, also confirms the absence of cracks in most of the intended regions of the sample treated using HPT followed by SPS. However, cracks are observed in all indented regions for the sample treated using only SPS. Additionally, EDS analyses do not show any difference at the microscopic scale between the distribution of elements for the two samples in both indented and non-indented regions.

The average of indentation diagonal size ($a$) and half crack length from the indentation tip to the crack tip ($c$) in Fig. 6 are measured as 35.2 µm and 2.4 µm for the HPT+SPS-treated sample, and 35.8 µm and 26.7 µm for the SPS-treated sample. From these values, the fracture toughness can be roughly estimated using the Anstis equation [40].

$K_{IC} = 0.015 \; (E/H)^{1/2} \; P \; c^{-3/2}$ (1)

where $K_{IC}$ is the fracture toughness (Pa·m$^{1/2}$), $E$ is the elastic modulus (Pa), which is about 220 GPa for YSZ [41], $H$ is the Vickers hardness (Pa), $P$ is the applied load (N) and $c$ is the half-crack length (m). This equation leads to values of $K_{IC}$ = 47.4 MPa·m$^{1/2}$ and 1.6 MPa·m$^{1/2}$ for the samples treated by HPT+SPS and SPS, respectively. The value of $K_{IC}$ = 47.4 MPa·m$^{1/2}$ is comparable to the fracture toughness of metallic materials, although this number should be evaluated with care as it was estimated locally by the indentation test and not by a standard fracture toughness experiment. The high toughness for the HPT+SPS-treated sample is likely due to the presence of dislocations, as discussed below [12,13].

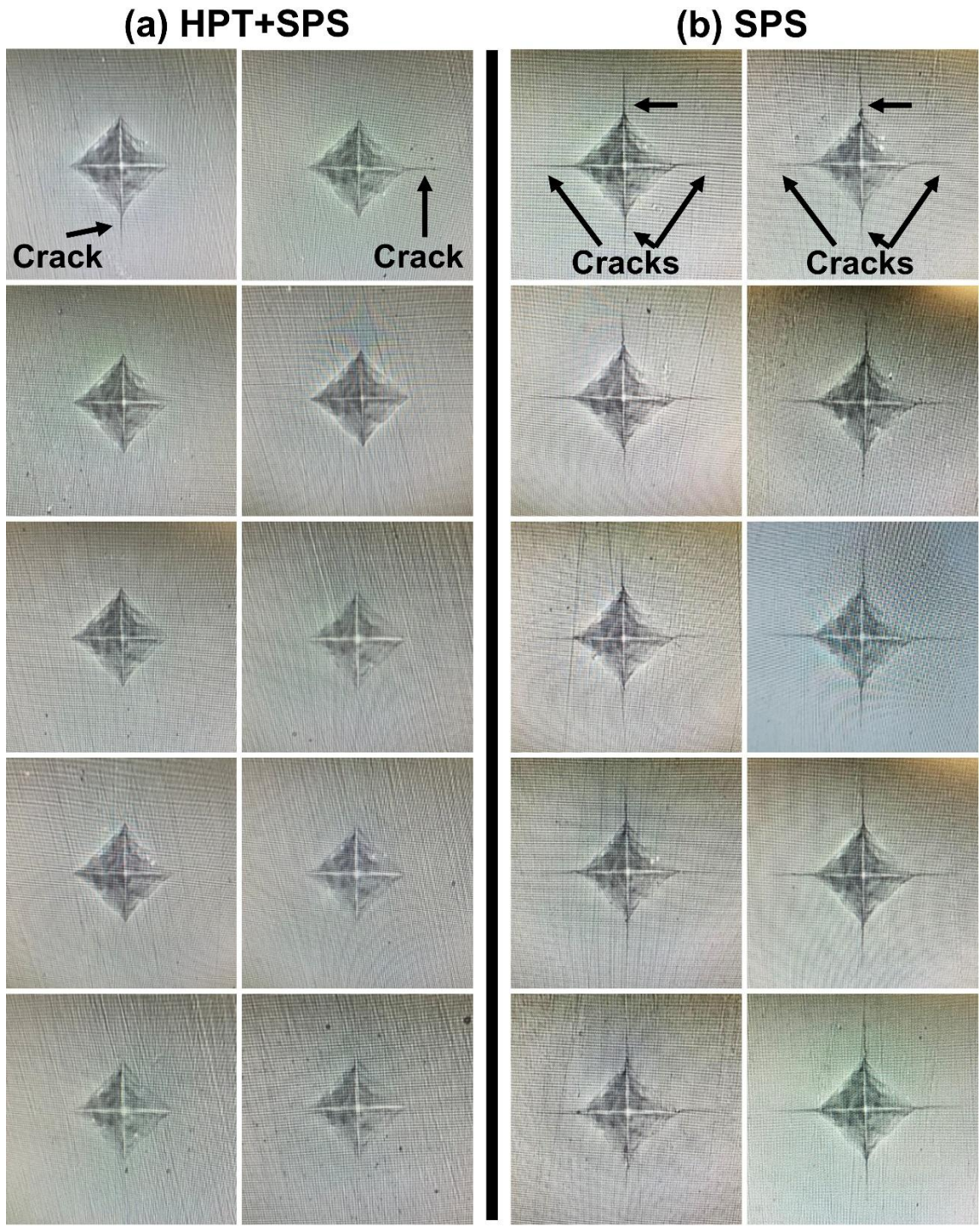

Figure 6. Images of indentation regions using a load of 1 kgf for yttria-stabilized zirconia samples consolidated by (a) high-pressure torsion followed by spark plasma sintering (HPT+SPS) and (b) only spark plasma sintering (SPS).

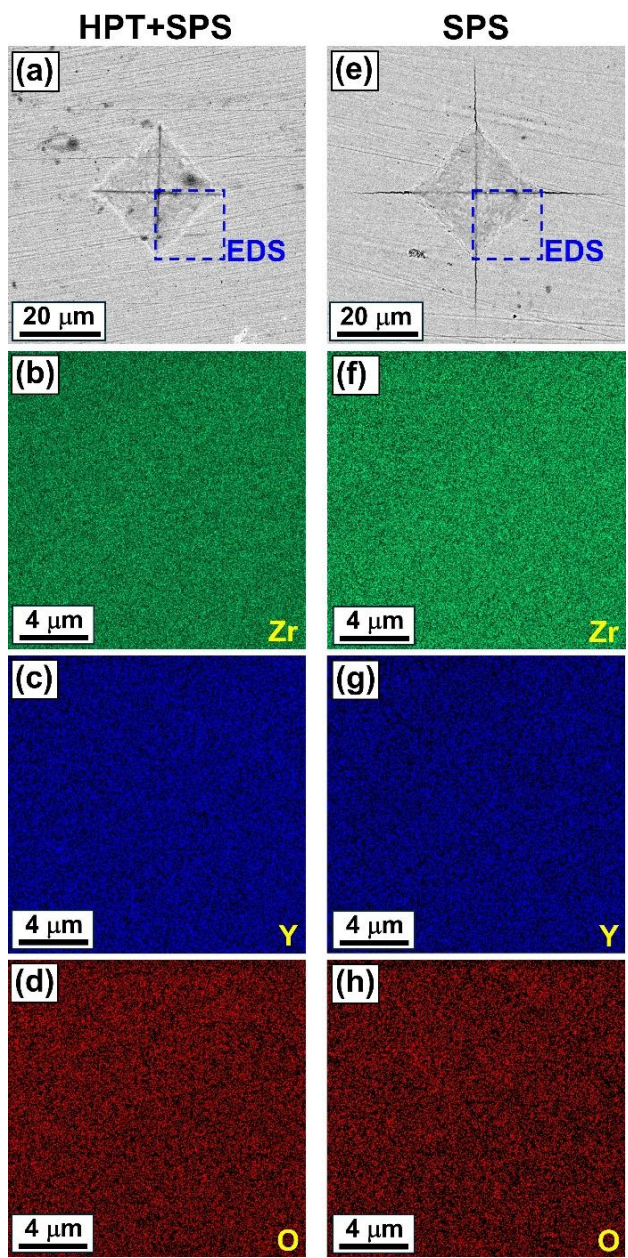

Figure 7. (a, e) SEM micrographs of indentation regions using a load of 1 kgf and (b-d, f-h) EDS elemental mappings from squared regions indicated in SEM micrographs for yttria-stabilized zirconia samples consolidated by (a-d) high-pressure torsion followed by spark plasma sintering (HPT+SPS) and (e-h) only spark plasma sintering (SPS).

## 4. Discussion

The present study on the simultaneous achievement of high hardness, high density and good toughness in YSZ raises two key questions regarding the mechanical properties: (i) how do the hardness and density compare with literature-reported values for bulk zirconia, and (ii) what is the underlying mechanism liable for the observed synergy of high microhardness and high toughness?

To address the first question, a direct comparison with reported data in the literature shows that the combination of HPT and SPS results in some improvement in mechanical properties. The hardness of the HPT+SPS-processed sample reaches 1500 ± 70 Hv, which is notably larger compared to the typical range of 1150-1450 Hv reported for conventionally sintered $ZrO_2$ - 3 mol% $Y_2O_3$-stabilized zirconia [2,38]. This enhancement in hardness is achieved while maintaining a density of 6.07 g/cm$^3$, which is very near to the nominal density of $ZrO_2$ - 3 mol% $Y_2O_3$ (6.13 g/cm$^3$ [37]). Unlike many nanocrystalline ceramics that suffer from porosity-related degradation in mechanical properties, the near-theoretical density of the current material confirms the effectiveness of HPT+SPS in promoting densification. To further illustrate this improvement, Fig. 8a presents a comparison of hardness and density values of various zirconia materials reported in the literature, including conventionally sintered $ZrO_2$ - 3 mol% $Y_2O_3$, SPS-processed samples, and other advanced processing routes [39,42-46]. As illustrated in Fig. 8a, the HPT+SPS-treated sample exhibits a unique combination of high hardness and near-theoretical density, positioning it among the highest-performing zirconia ceramics. In contrast, many conventionally sintered samples show either lower hardness due to grain coarsening or reduced density due to residual porosity. Although density plays an important role in the hardness of ceramics, the effect of small grain sizes on the hardness through the Hall-Petch relationship is also of significance in fully dense samples [39,47]. A Hall-Petch plot is shown in Fig. 8b for the zirconia ceramics with different grain sizes ($d$ in nm) reported in the literature [39,43,45,46] and in the current study, including two equations given for the hardness ($H$ in Hv) of YSZ with cubic and tetragonal structures by Trunec [39] and Novitskaya *et al.* [47], respectively.

$H = 1110 + 3520d^{-1/2}$ (2)

$H = 1059 + 3280d^{-1/2}$ (3)

While many datum points scatter from these Hall-Petch relationships due to the effect of other factors such as porosity, the HPT+SPS-processed sample follows the relationship due to its high relative density. The slightly higher hardness of the HPT+SPS-processed sample compared to the equation for the tetragonal zirconia [47] can be due to the presence of other strengthening mechanisms such as dislocations [48]. The observed trends in Fig. 8 reinforce the advantage of combining severe plastic deformation for cold pre-consolidation [15-18] with SPS for final rapid consolidation [25-27,35], demonstrating that this approach enables significant strengthening without compromising densification and grain size, an issue that is of importance in many applications of YSZ [1-4].

The second question concerns the origin of the high toughness ($K_{IC}$ = 47.4 MPa·m$^{1/2}$, estimated from the indentation test) in the HPT+SPS-processed sample, despite its high hardness. Indentation testing reveals that, unlike the sample processed only by SPS, which forms cracks at all indentation tips, the HPT+SPS-processed sample exhibits minimal crack formation, suggesting enhanced resistance to crack propagation. This is a remarkable observation because achieving high toughness in nanostructured ceramics remains a major challenge [11]. One plausible explanation is the high density of dislocations within the nanograins, introduced by severe plastic deformation during HPT [15-18]. While dislocations are typically rare in ceramics due to their strong ionic and/or covalent bonds, previous works have found that HPT can generate a very high dislocation density in certain ceramics [19,33]. In the current material, these dislocations, which partly survive during SPS, likely serve as effective mechanisms for stress redistribution, thereby increasing toughness by allowing localized plasticity [12,13]. Another contributing factor is the refined grain size of 80 nm, which not only enhances hardness through the Hall-Petch effect [8-10] but also plays a role in toughening in the presence of dislocations [48]. In conventional YSZ, toughness is primarily ascribed to stress-induced phase transition from tetragonal or cubic to monoclinic, which leads to volumetric expansion and crack deflection [5]. However, in the present case, the toughening mechanism is likely dominated by dislocation-mediated plasticity [12,13], where the interaction between dislocations and nanograins helps dissipate energy and reduce crack propagation [48].

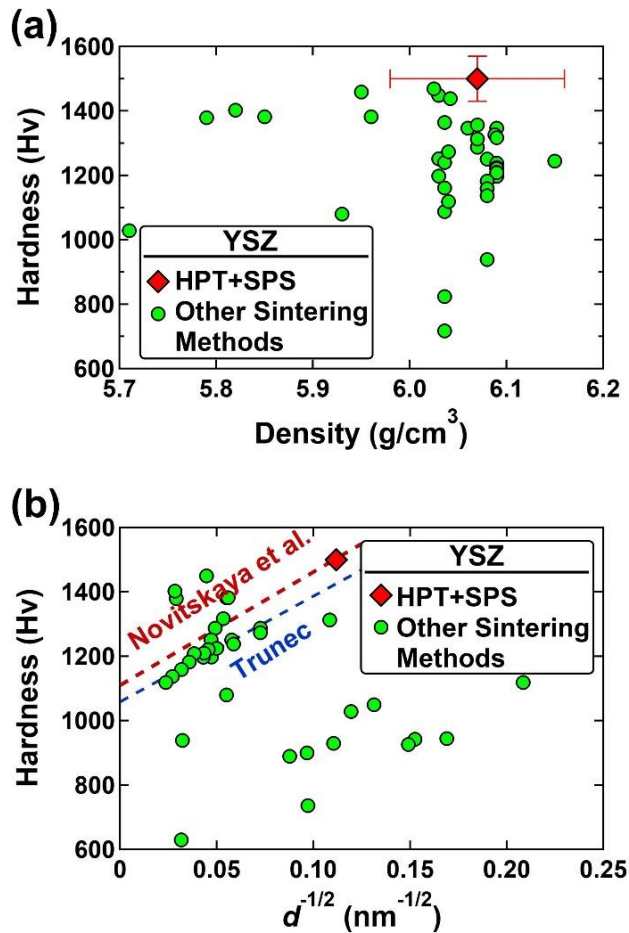

Figure 8. Variation of hardness versus (a) density and (b) inverse root of grain size ($d^{-1/2}$) for zirconia bulk samples reported in literature [39,42-47] compared to yttria-stabilized zirconia sintered using high-pressure torsion followed by spark plasma sintering (HPT+SPS) in this study. Hall-Petch relationships reported by Trunec [39] and Novitskaya *et al.* [47] were included in (b) using dotted lines.

The observed mechanical behavior suggests that the combination of nanograin refinement and dislocation engineering offers a novel pathway to achieving both high hardness and toughness in zirconia ceramics. Unlike conventional sintering, which typically results in coarse-grained microstructures with limited dislocation activity, the HPT+SPS approach allows for a unique interplay between grain boundaries, dislocations and densification. The synergy between these factors overcomes the traditional trade-off between hardness and toughness in ceramics, opening new possibilities for optimizing mechanical properties through advanced severe plastic deformation processing techniques. Since various kinds of ceramics have already been processed by HPT [49-51], these findings also have potential implications for other ceramic systems, where severe plastic deformation combined with rapid sintering may serve as a promising strategy for enhancing mechanical performance.

## 5. Conclusions

This study demonstrates the successful production of bulk nanostructured yttria-stabilized zirconia by integrating high-pressure torsion (HPT) and spark plasma sintering (SPS). The processed material achieves a high hardness of 1500 Hv, exceeding the typical hardness range of conventionally sintered $ZrO_2$ - 3 mol% $Y_2O_3$, while maintaining 6.07 g/cm³ density, close to the theoretical density. The indentation results confirm enhanced fracture toughness ($K_{IC}$ = 47.4 MPa·m$^{1/2}$, estimated from the indentation test) as the HPT+SPS-processed sample exhibits minimal crack formation compared to the sample processed only by SPS. The combination of nanograin refinement and a high density of dislocations plays a crucial role in achieving both high hardness and toughness, offering an alternative strain-induced dislocation-based toughening mechanism that does not rely on the metastable-to-stable phase transition. These findings highlight the potential of combining severe plastic deformation and rapid sintering as an effective approach for designing advanced ceramic materials with superior mechanical properties.

**CRediT Authorship Contribution Statement**

All authors: Conceptualization, Methodology, Investigation, Validation, Writing – review & editing.

**Declaration of competing interest**

The authors declare no competing interests that could have influenced the results reported in the current article.

**Data availability**

All the data presented in this article are made available through the request to the corresponding author

**Acknowledgments**

The author S.D. acknowledges the Ministry of Education, Culture, Sports, Science and Technology of Japan for a scholarship. This research received support partly from the Japan Society for the Promotion of Science (grant number: JP22K18737).